# Fluctuation theorem for the renormalized entropy change in the strongly nonlinear nonequilibrium regime

Yuki Sughiyama[1] and Sumiyoshi Abe[2,3]

[1]*Institute of Physics, University of Tsukuba, Ibaraki 305-8571, Japan*

[2]*Department of Physical Engineering, Mie University, Mie 514-8507, Japan*

[3]*Institut Supérieur des Matériaux et Mécaniques Avancés, 44 F. A. Bartholdi,*

*72000 Le Mans, France*

**Abstract**   Generalizing a recent work [T. Taniguchi and E. G. D. Cohen, J. Stat. Phys. **126**, 1 (2006)] that was based on the Onsager-Machlup theory, a nonlinear relaxation process is considered for a macroscopic thermodynamic quantity. It is found that the fluctuation theorem holds in the nonlinear nonequilibrium regime if the change of the entropy characterized by local equilibria is appropriately renormalized. The fluctuation theorem for the ordinary entropy change is recovered in the linear near-equilibrium case. This result suggests a possibility that the information-theoretic entropy of the Shannon form may be modified in the strongly nonlinear nonequilibrium regime.

PACS number(s): 05.70.Ln, 05.40.-a, 05.10.Gg

## I. INTRODUCTION

Thermodynamics is concerned with the averages of macroscopic physical quantities, whereas equilibrium statistical mechanics can give information on fluctuations around the averages. Einstein's 1910 theory of fluctuations [1,2] builds a bridge between the two based on the thermodynamic entropy and the reversal of Boltzmann's relation for it. In nonequilibrium, fluctuations are considered to play a vital role and may cause a richer variety of phenomena than those in equilibrium. In spite of a lot of efforts, it seems fair to say that simple and complete laws could not have been found for describing universal physical properties of fluctuations in system states far from equilibrium. However, the situation has changed when the so-called fluctuation theorem was formulated in the middle of 1990's [3,4].

The fluctuation theorem reveals a kind of symmetry hidden behind the distributions of the entropy change, quantity of heat, work and so on in nonequilibrium situations. It holds for stochastic systems [5,6] as well as deterministic chaos [4]. Also, several real experiments have been performed, and good agreements of the predictions of the theorem with the experimental results have been reported [7-9].

In a recent work [10], the fluctuation theorem has been rederived by making use of the Onsager-Machlup theory [11], in which the existence of local equilibria and linearity of relaxation process are essential premises. Discussions of this kind are of obvious importance, since a macroscopic theory (i.e., thermodynamics) plays a guiding role for consistently developing a microscopic approach (i.e., statistical mechanics). The authors of Ref. [10] apply a constant external dragging force and consider the fluctuations of work, friction, and quantity of heat. A problem of crucial importance here is that the linear approximation is not legitimate in the strongly nonequilibrium regime. Accordingly, the external force cannot be so strong either.

In this paper, we generalize the discussion in Ref. [10] to a nonlinear case and show that the transient fluctuation theorem still holds if the change of the entropy characterized by local equilibria is appropriately modified. This modified quantity is referred to as the "renormalized entropy change". Our result suggests a possibility that the ordinary information-theoretic entropy of the Shannon form may be modified in the strongly nonlinear nonequilibrium regime.

## II. NONLINEAR NONEQUILIBRIUM PROCESS: GENERALIZATION OF THE ONSAGER-MACHLUP THEORY

To be self-contained, this section is devoted to the preparation for our main discussion in Sec. III.

Suppose that the total system consists of the objective system and the surrounding environment and is initially not in equilibrium. Consider the evolution of a macroscopic physical quantity of the objective system, the energy $\phi$ here, along a process from a given arbitrary initial state to a certain nonequilibrium stationary state. We formulate the dynamics of $\phi$ by employing the Langevin equation

$$\frac{d\phi}{dt} = F(\phi) + \xi. \tag{1}$$

Here, $F(\phi)$ is a current and $\xi$ is the Gaussian white noise satisfying

$$\overline{\xi(t)} = 0, \qquad \overline{\xi(t)\xi(t')} = 2D\delta(t-t'), \tag{2}$$

where the over-bars stand for the averages over the noise distribution.

Onsager and Machlup [11] assume the initial state of the objective system to be close to equilibrium and discuss its relaxation to equilibrium. Accordingly, they are able to express the current in terms of the thermodynamic force, $dS^{tot}(\phi)/d\phi$, as $F(\phi) = L\, dS^{tot}(\phi)/d\phi$, where $S^{tot}(\phi)$ and $L$ are the total entropy and the transport coefficient, respectively. Since the total system is in a state near equilibrium, its entropy can well be approximated by a quadratic function [1,2]: $S^{tot}(\phi) = \text{const.} - (1/2)\alpha\phi^2$ with $\alpha > 0$ (here, $\phi_0$ yielding $S^{tot}(\phi_0) = \max$ is taken to be zero for the sake of simplicity). Accordingly, the distribution of fluctuations, $\phi$, is Gaussian: $\rho_\infty(\phi) \propto \exp[S^{tot}(\phi)] \sim \exp[-(1/2)\alpha\phi^2]$ (Boltzmann's constant being set equal to unity), and the Langevin equation in Eq. (1) becomes linear. To realize a nonequilibrium stationary state, the authors of Ref. [10] introduce a constant external dragging force.

A point here is that if the entropy is a quadratic function and the external force is not applied, then the physics is well determined in the neighborhood of $\phi = \phi_0 (= 0)$.

However, for a far-from-equilibrium system exhibiting slow relaxation, the entropy is not quadratic and may have a complex landscape with a number of local maxima, in general. Therefore, in such cases, the state of the system can still be characterized around local maxima even without external dragging forces. Thus, we renounce the linearities of both the current and the thermodynamics force with respect to $\phi$ and do not apply external dragging forces.

To find one such nonequilibrium stationary state, we consider the following Fokker-Planck equation for the probability distribution $\rho(\phi, t) = \overline{\delta(\phi - \phi_\xi(t))}$:

$$\frac{\partial \rho(\phi, t)}{\partial t} = -\frac{\partial}{\partial \phi}[F(\phi)\rho(\phi, t)] + D\frac{\partial^2 \rho(\phi, t)}{\partial \phi^2}, \qquad (3)$$

where $\phi_\xi(t)$ is the solution of Eq. (1). A stationary solution of this equation is given by

$$\rho_S(\phi) \propto e^{\Sigma(\phi)}, \qquad (4)$$

where $\Sigma(\phi)$ is connected to $F(\phi)$ as follows:

$$F(\phi) = D\frac{d\Sigma(\phi)}{d\phi}. \qquad (5)$$

Later, we shall see how $\Sigma(\phi)$ is related to the renormalized entropy change.

Take a time interval $[0, \tau]$ and impose the boundary conditions, $\phi(0) = X$ and $\phi(\tau) = Y$. The forward transition probability from $X$ to $Y$ is given by the following functional integral [12,13]:

$$f_F(Y, \tau \mid X, 0) = N \int D\xi \int_{\phi(0)=X}^{\phi(\tau)=Y} D\phi \, \delta[\phi - \phi_\xi] \exp\left[-\frac{1}{4D}\int dt \, \xi^2(t)\right]$$

$$= N \int_{\phi(0)=X}^{\phi(\tau)=Y} D\phi \, \text{Det}\left[\left(\frac{d}{dt} - \frac{dF(\phi)}{d\phi}\right)\delta(t-t')\right]$$

$$\times \exp\left[-\frac{1}{4D}\int_0^\tau dt \left(\frac{d\phi}{dt} - \frac{dF(\phi)}{d\phi}\right)^2\right], \quad (6)$$

where the subscript "$F$" denotes the forward process [6], $N$ is a normalization factor that will commonly be used throughout this paper, $\delta[\phi - \phi_\xi] \equiv \prod_t \delta(\phi(t) - \phi_\xi(t))$, and the functional determinant is defined for the continuous indices $t$ and $t'$.

To evaluate the determinant, we employ the standard manipulation [13]: $\text{Det } M = \exp(\text{Tr ln } M)$, where

$$M(t, t') \equiv \left(\frac{d}{dt} - \frac{dF(\phi)}{d\phi}\right)\delta(t-t'). \quad (7)$$

Write the matrix as follows:

$$M(t, t') = \left(\frac{d}{dt}\right) K(t, t'), \quad (8)$$

$$K(t, t') = \delta(t-t') - \theta(t-t') \frac{dF(\phi(t'))}{d\phi(t')}, \quad (9)$$

where $\theta(x)$ is the Heaviside step function. Notice that we are using the "forward propagator", $\theta(t-t')$, in Eq. (9). Absorbing $\exp[\text{Tr}\ln(d/dt)]$ in the normalization factor and expanding the logarithm, we have

$$\ln(\text{Det}\, K) = -\theta(0)\int_0^\tau dt\, \frac{dF(\phi)}{d\phi}$$
$$-\frac{1}{2}\int_0^\tau dt_1 \int_0^\tau dt_2\, \theta(t_1-t_2)\theta(t_2-t_1)\frac{dF(\phi(t_1))}{d\phi(t_1)}\frac{dF(\phi(t_2))}{d\phi(t_2)}$$
$$-\cdots\cdots. \tag{10}$$

In this series, only the first term survives because the products of the step functions vanish. Setting $\theta(0) = 1/2$, we obtain

$$\text{Det}\, M \propto \exp\left[-\frac{1}{2}\int_0^\tau dt\, \frac{dF(\phi)}{d\phi}\right]. \tag{11}$$

Consequently, the forward transition probability is expressed as follows:

$$f_F(Y,\tau|X,0) = N \int_{\phi(0)=X}^{\phi(\tau)=Y} D\phi \exp\left(-\int_0^\tau dt\, L\right), \tag{12}$$

where

$$L = \frac{1}{4D}\left(\frac{d\phi}{dt} - F(\phi)\right)^2 + \frac{1}{2}\frac{dF(\phi)}{d\phi} \tag{13}$$

is the "thermodynamic Lagrangian". The second term on the right-hand side highlights an effect of the nonlinearity.

## III. FLUCTUATION THEOREM FOR THE RENORMALIZED ENTROPY CHANGE

With the preparation in the preceding section, now we are in a position to discuss a transient fluctuation theorem in a process from a given arbitrary initial state, $\rho(X,0)$, to the nonequilibrium stationary state, i.e., $\rho_S(\phi)$ in Eq. (4). To find a relevant physical quantity, first we consider the time reversal operation: $t = -\tilde{t}$. $\phi$ is assumed to transform as a scalar variable: $\phi(t) = \tilde{\phi}(\tilde{t})$. Under this operation, the thermodynamic Lagrangian transforms as

$$L\left(d\phi(t)/dt, \phi(t)\right) = L\left(d\tilde{\phi}(\tilde{t})/d\tilde{t}, \tilde{\phi}(\tilde{t})\right) + \frac{1}{D} F(\tilde{\phi}(\tilde{t})) \frac{d\tilde{\phi}(\tilde{t})}{d\tilde{t}}. \tag{14}$$

Quite remarkably, the second term on the right-hand side is the $\tilde{t}$-derivative of $\Sigma(\tilde{\phi}(\tilde{t}))$ with $\Sigma$ appearing in Eq. (5). That is,

$$L\left(d\phi(t)/dt, \phi(t)\right) = L\left(d\tilde{\phi}(\tilde{t})/d\tilde{t}, \tilde{\phi}(\tilde{t})\right) + \frac{d\Sigma(\tilde{\phi}(\tilde{t}))}{d\tilde{t}}. \tag{15}$$

Accordingly, the transition probability changes as follows:

$$f_F(Y, \tau | X, 0) = N\, e^{\Sigma(Y) - \Sigma(X)} \int_{\tilde{\phi}(0) = X}^{\tilde{\phi}(-\tau) = Y} D\tilde{\phi} \exp\left[-\int_{-\tau}^{0} d\tilde{t}\, L\left(d\tilde{\phi}(\tilde{t})/d\tilde{t}, \tilde{\phi}(\tilde{t})\right)\right]. \tag{16}$$

Doing the shift, $\hat{t} = \tilde{t} + \tau$, and noticing $\tilde{\phi}(\tilde{t}) = \hat{\phi}(\hat{t})$ as well as the invariance of the functional integral part under time translation, we obtain

$$f_F(Y, \tau | X, 0) \rho_S(X) = f_F(X, \tau | Y, 0) \rho_S(Y), \tag{17}$$

where $\rho_S$ is the nonequilibrium stationary state in Eq. (4). Thus, in the present nonlinear nonequilibrium system, holds the detailed balance condition, which is regarded as a remnant of microscopic reversibility [14]. Notice, however, that the quantities treated here are the macroscopic thermodynamic variables.

It is also noticed that if the total derivative term is not extracted in Eq. (14), one obtains the reverse transition probability $f_R(X, \tau | Y, 0)$, which is related to the forward transition probability as follows:

$$\begin{aligned} f_F(Y, \tau | X, 0) &= f_R(X, \tau | Y, 0) \\ &= N \int_{\hat{\phi}(\tau)=X}^{\hat{\phi}(0)=Y} D\hat{\phi} \exp\left(-\int_0^\tau d\hat{t}\, \hat{L}\right), \end{aligned} \tag{18}$$

where

$$\hat{L} = \frac{1}{4D}\left(-\frac{d\hat{\phi}}{d\hat{t}} - F(\hat{\phi})\right)^2 + \frac{1}{2}\frac{dF(\hat{\phi})}{d\hat{\phi}} \tag{19}$$

Now, the proof of the fluctuation theorem is straightforward. The quantity to be considered is $\Sigma(\phi)$, as suggested by the structure in Eq. (17). So, let us evaluate in the following way the probability that the amount of its change along a process from a given arbitrary initial state, $\rho(X, 0)$, to a nonequilibrium stationary state, $\rho_S(\phi)$, during the time interval $[0, \tau]$ is $\Delta\Sigma$:

$$P_F(\Delta\Sigma) = \left\langle \delta\left(\Delta\Sigma - \int_0^\tau dt\, \frac{d\Sigma(\phi)}{dt}\right)\right\rangle_F$$

$$\equiv \iint dX\, dY\, \delta\left(\Delta\Sigma - [\Sigma(Y) - \Sigma(X)]\right) f_F(Y, \tau | X, 0)\, \rho(X, 0). \tag{20}$$

From the detailed balance condition in Eq. (17), we have

$$P_F(\Delta\Sigma) = e^{\Delta\Sigma} \iint dX\, dY\, \delta\left(\Delta\Sigma - [\Sigma(Y) - \Sigma(X)]\right) f_F(X, \tau | Y, 0)\, \rho(X, 0). \tag{21}$$

Interchanging the integration variables, $X$ and $Y$, and using Eq. (18), we find

$$P_F(\Delta\Sigma) = e^{\Delta\Sigma} \iint dX\, dY\, \delta\left(-\Delta\Sigma - [\Sigma(Y) - \Sigma(X)]\right) f_F(Y, \tau | X, 0)\, \rho(Y, 0)$$

$$= e^{\Delta\Sigma} \iint dX\, dY\, \delta\left(-\Delta\Sigma - [\Sigma(Y) - \Sigma(X)]\right) f_R(X, \tau | Y, 0)\, \rho(Y, 0)$$

$$= e^{\Delta\Sigma} \left\langle \delta\left(-\Delta\Sigma - \int_0^\tau dt\, \frac{d\Sigma(\phi(t))}{dt}\right)\right\rangle_R$$

$$= e^{\Delta\Sigma} P_R(-\Delta\Sigma). \tag{22}$$

where the subscript "$R$" indicates the reverse process [6]. Therefore, we obtain

$$\frac{P_F(\Delta\Sigma)}{P_R(-\Delta\Sigma)} = e^{\Delta\Sigma}, \tag{23}$$

which is the main result of the present work.

A remaining task is to elucidate the physical meaning of the quantity, $\Delta\Sigma$. Clearly, it is not the change of the ordinary entropy, $S^{tot}(\phi)$, defined by local equilibria. From Eq. (5), we find that

$$\frac{d\Sigma(\phi)}{dt} = \frac{1}{D}F(\phi)\frac{d\phi}{dt}$$
$$= \frac{1}{D}\frac{\chi(\Gamma)}{\Gamma}\frac{dS^{tot}(\phi)}{dt}, \qquad (24)$$

where $\chi(\Gamma) \equiv F(\phi)$, in which $\phi$ is solved in terms of the affinity [15] defined by

$$\Gamma = \frac{dS^{tot}(\phi)}{d\phi}. \qquad (25)$$

If the total entropy is given by the sum of the entropies of the objective and environmental systems, then $\Gamma$ is given by the difference between the inverse temperatures of these subsystems. Equation (24) puts a basis for calling $\Delta\Sigma$ the renormalized entropy change.

In a particular case when the total system is in a state near equilibrium, the linear approximation is valid well: that is, $\chi(\Gamma) = L\Gamma$ with the transport coefficient, $L$, satisfying the fluctuation-dissipation theorem, $D = L$. Then, Eq. (23) becomes reduced to the transient fluctuation theorem for the entropy change, which is known in the literature (see Ref. [16], for example).

## V. CONCLUSION

We have examined the fluctuation theorem by generalizing the discussion in Ref. [10] to the case of a nonlinear slow relaxation process for the macroscopic thermodynamic energy. In this way, a system in a state far from equilibrium is consistently treated. We have found that the transient fluctuation theorem holds if the entropy change is appropriately renormalized. This result suggests a possibility that Shannon's definition of the entropy, $S = -\sum_i p_i \ln p_i$ (with a probability distribution $\{p_i\}$), which is connected to $S^{tot}$ through the maximum entropy principle [17], may be modified in the strongly nonequilibrium regime.

## ACKNOWLEDGMENTS

Y. S. would like to thank the Department of Physical Engineering, Mie University, for the hospitality extended to him. He is also indebted to the Yukawa Institute for Theoretical Physics at Kyoto University, where this work was initiated during the YITP-W-07-07 on "Thermal Quantum Field Theory and Its Applications". S. A. thanks Eddie Cohen for discussions. This work was supported in part by Grant-in-Aid for Scientific Research (B) of the Ministry of Education.

## REFERENCES

[1]  A. Einstein, Ann. der Phys. **33**, 1275 (1910).

[2] L. D. Landau and E. M. Lifshitz, *Statistical Physics* 3rd edition (Pergamon Press, Oxford, 1980).

[3] D. J. Evans, E. G. D. Cohen, and G. P. Morriss, Phys. Rev. Lett. **71**, 2401 (1993): Erratum **71**, 3616 (1993).

[4] G. Gallavotti and E. G. D. Cohen, Phys. Rev. Lett. **74**, 2694 (1995).

[5] J. L. Lebowitz and H. Spohn, J. Stat. Phys. **95**, 333 (1999).

[6] G. E. Crooks, Phys. Rev. E **60**, 2721 (1999); ibid. E **61**, 2361 (2000).

[7] G. M. Wang, E. M. Sevick, E. Mittag, D. J. Searles, and D. J. Evans, Phys. Rev. Lett. **89**, 050601 (2002).

[8] S. Ciliberto, N. Garnier, S. Hernandez, C. Lacpatia, J.-F. Pinton, G. R. Chavarria, Physica A **340**, 240 (2004).

[9] N. Garnier and S. Ciliberto, Phys. Rev. **71**, 060101(R) (2005).

[10] T. Taniguchi and E. G. D. Cohen, J. Stat. Phys. **126**, 1 (2006).

[11] L. Onsager and S. Machlup, Phys. Rev. **91**, 1505 (1953).

[12] E. Gozzi, Phys. Rev. D **30**, 1218 (1984).

[13] J. Zinn-Justin, *Quantum Field Theory and Critical Phenomena*, 4th Ed. (Oxford University Press, Oxford, 2002).

[14] R. C. Tolman, *The Principles of Statistical Mechanics* (Dover, New York, 1979).

[15] M. Le Bellac, F. Mortessagne, and G. G. Batrouni, *Equilibrium and Non-Equilibrium Statistical Thermodynamics* (Cambridge University Press,

Cambridge, 2004).

[16] U. Seifert, Phys. Rev. Lett. **95**, 040602 (2005).

[17] *E. T. Jaynes: Papers on Probability, Statistics and Statistical Physics*, edited by R. D. Rosenkrantz (Kluwer, Dordrecht, 1989).